\documentclass[article, twocolumn, superscriptaddress]{revtex4}

\usepackage{graphicx}
\usepackage{amsfonts}

\begin{document}
\title{Single photon emitters in exfoliated WSe$_2$ structures.}
\author{M. Koperski}
\affiliation{Laboratoire National des Champs Magn\'{e}tiques Intenses, CNRS-UJF-UPS-INSA, 25 Rue des Martyrs, 38042 Grenoble, France}
\affiliation{Institute of Experimental Physics, Faculty of Physics, University of Warsaw, Ho\.{z}a 69, 00-681 Warsaw, Poland}
\author{K. Nogajewski}
\affiliation{Laboratoire National des Champs Magn\'{e}tiques Intenses, CNRS-UJF-UPS-INSA, 25 Rue des Martyrs, 38042 Grenoble, France}
\author{A. Arora}
\affiliation{Laboratoire National des Champs Magn\'{e}tiques Intenses, CNRS-UJF-UPS-INSA, 25 Rue des Martyrs, 38042 Grenoble, France}
\author{J. Marcus}
\affiliation{Institut N\'eel, CNRS-UJF, BP 166, 38042 Grenoble, France}
\author{P. Kossacki}
\affiliation{Laboratoire National des Champs Magn\'{e}tiques Intenses, CNRS-UJF-UPS-INSA, 25 Rue des Martyrs, 38042 Grenoble, France}
\affiliation{Institute of Experimental Physics, Faculty of Physics, University of Warsaw, Ho\.{z}a 69, 00-681 Warsaw, Poland}
\author{M. Potemski}
\affiliation{Laboratoire National des Champs Magn\'{e}tiques Intenses, CNRS-UJF-UPS-INSA, 25 Rue des Martyrs, 38042 Grenoble, France}

\maketitle \textbf{Crystal structure imperfections in solids often
act as efficient carrier trapping centers which, when suitably
isolated, act as sources of single photon emission. The best known
examples of such attractive imperfections are well-width or
composition fluctuations in semiconductor heterostructures
\cite{reed1988, michler2000} (resulting in a formation of quantum
dots) and coloured centers in wide bandgap (e.~g., diamond)
materials \cite{gruber1997,brouri2000,mizuochi2012}. In the case
of recently investigated thin films of layered compounds, the
crystal imperfections may logically be expected to appear at the
edges of commonly investigated few-layer flakes of these
materials, exfoliated on alien substrates. Here, we report on
comprehensive optical micro-spectroscopy studies of thin layers of
tungsten diselenide, WSe$_2$, a representative semiconducting
dichalcogenide with a bandgap in the visible spectral range. At
the edges of WSe$_2$ flakes, transferred onto Si/SiO$_2$
substrates, we discover centers which, at low temperatures, give
rise to sharp emission lines (100~$\mu$eV linewidth). These narrow
emission lines reveal the effect of photon antibunching, the
unambiguous attribute of single photon emitters. The optical
response of these emitters is inherently linked to two-dimensional
properties of the WSe$_2$ monolayer, as they both give rise to
luminescence in the same energy range, have nearly identical
excitation spectra and very similar, characteristically large
Zeeman effects. With advances in the structural control of edge
imperfections, thin films of WSe$_2$ may provide added
functionalities, relevant for the domain of quantum
optoelectronics. }

\begin{figure}
\includegraphics[width=8.5cm]{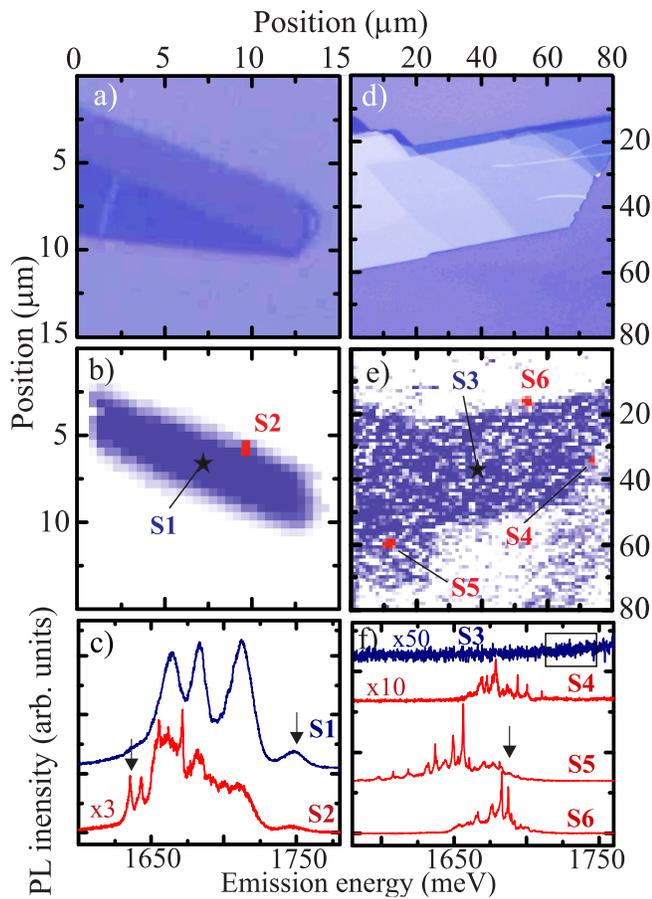}
\caption{\textbf{Images and scanning $\mu$PL spectroscopy on
WSe$_2$ layers, revealing narrow line emission centers at the
edges of flakes deposited on Si/SiO$_2$ substrates.} \textbf{a},
Optical microscope images of a thin WSe$_2$ flake. A large part of
this flake consists of a WSe$_2$ monolayer whose image is
reproduced in \textbf{b}, with a contour plot (violet contrast) of
the intensity of the $\mu$PL detected at the selected emission
energy of E$\sim$1.75 eV (free-exciton resonance of a WSe$_2$
monolayer). \textbf{c}, $\mu$PL spectra measured from two selected
spots of the WSe$_2$ monolayer, S1 and S2. The PL at the central
spot (S1) is a known spectrum of the WSe$_2$ monolayer and indeed
characteristic of the majority of spots on the surface of our
monolayer. Sharp emission lines are seen on the top of the rather
broad S1-like spectrum at specific spots (e. g., S2,) located at
the edges of the flake. The location of S2 is shown in \textbf{b},
with a contour plot (red contrast) of the intensity of the sharp
line observed at E=1.636 eV. The thick flake of WSe$_2$
illustrated with the optical-microscope image in \textbf{d}, shows
a suppressed PL but nevertheless its image can be reproduced in
the course of $\mu$PL scanning experiments as shown by the contour
plot with violet contrast in \textbf{e}, which represents the
intensity map of scattered light detected in the energy interval
1.71-1.75 eV. Shown in \textbf{f}, the narrow-line emission
spectra, S4-S5, are found at specific edge locations of the thick
WSe$_2$ flake, as illustrated in \textbf{e} with the contour plot
(red contrast) of the PL monitored at E= 1.687 eV.}
\label{fig:map}
\end{figure}

Investigations of thin layers of semiconducting transition metal
dichalcogenides (TMDs) are driven by scientific curiosity (to
uncover the properties of a new class of two-dimensional systems
with unconventional electronic bands) and by their possible new
functionalities arising from using a valley degree of freedom in
optoelectronic devices~\cite{wang2012}. Typically for the family
of semiconducting TMDs, a one-molecule-thick layer of WSe$_2$ is
known to display a robust photo-luminescence~\cite{zhao2012}. This
is due to its direct bandgap semiconductor structure in contrast
to the indirect bandgap structure of WSe$_2$ multilayers. Indeed,
when increasing the number of layers, the photoluminescence (PL)
of few-layer WSe$_2$ is progressively quenched and red shifted,
and becomes practically undetectable in sufficiently thick
layers~\cite{zhao2012}. Optical studies of WSe$_2$ layers are
routinely performed on flakes exfoliated from bulk material and
transferred onto Si/SiO$_2$ substrates. Such structures are also
the objects of this work.

Optical-microscope images of two selected WSe$_2$ flakes are shown
in Fig.~\ref{fig:map}a and \ref{fig:map}d. A large part of the
flake shown in Fig.~\ref{fig:map}a is identified as a monolayer,
attached to a thicker WSe$_2$ film. When this flake is scanned
with micro-PL spectra at low temperatures, the majority PL
response of the WSe$_2$ monolayer exibits a familiar form, known
from literature~\cite{jones2013}. As shown in Fig.~\ref{fig:map}c,
the emission spectrum of the WSe$_2$ monolayer is composed of
several, rather broad peaks ($\sim$20 meV linewidth). This
spectrum is characteristic of an n-type WSe$_2$
monolayer~\cite{jones2013} (unintentional doping in our case). The
upper energy peak (at $\sim$1.75 eV) is commonly attributed to a
direct, free exciton resonance; the subsequent lower energy peaks
are assigned to charged and bound (localized)
excitons~\cite{Wang2014}. The contour image of the WSe$_2$
monolayer is clearly reproduced with mapping the intensity of the
free exciton peak (see Fig.~\ref{fig:map}b). Interestingly enough,
there exist specific spots on this flake at which the broad PL
spectrum from the bulk of the WSe$_2$ monolayer tends to be
replaced by a series of sharp (100~$\mu$eV linewidth) emission
lines (see Fig.~\ref{fig:map}c). These narrow line emitting
centers (NLECs) are located at the edge of the WSe$_2$ flake. This
is shown with photoluminescence mapping of the intensity of one of
the well spectrally-resolved narrow emission lines (see
Fig.~\ref{fig:map}b). Markedly, the presence of NLECs is not only
a property of the WSe$_2$ monolayer. In
Fig.~\ref{fig:map}d~-~\ref{fig:map}f the optical microscope image,
the photoluminescence maps and spectra measured at selected spots
are presented for a thick WSe$_2$ flake. Its thickness varies
between 10 and 15 nm (6 - 9 monolayers), as estimated from AFM
measurements on its different parts. Here, we show the presence of
spots giving sharp emission lines, located again at the edges but,
this time, of a rather thick WSe$_2$ flake. The characteristic
NLECs have also been seen in a number of other WSe$_2$ flakes.
Every time the NLECs are found at the flake edges and sharp
emission lines always appear in the spectral range which overlaps
with the broad emission band of charged and localized excitons of
the WSe$_2$ monolayer. Studying the NLECs associated with thicker
WSe$_2$ flakes is advantageous as the emission from the bulk of
these flakes is suppressed and sharp edge-emission lines are
better resolved.

\begin{figure}
\includegraphics[width=8.5cm]{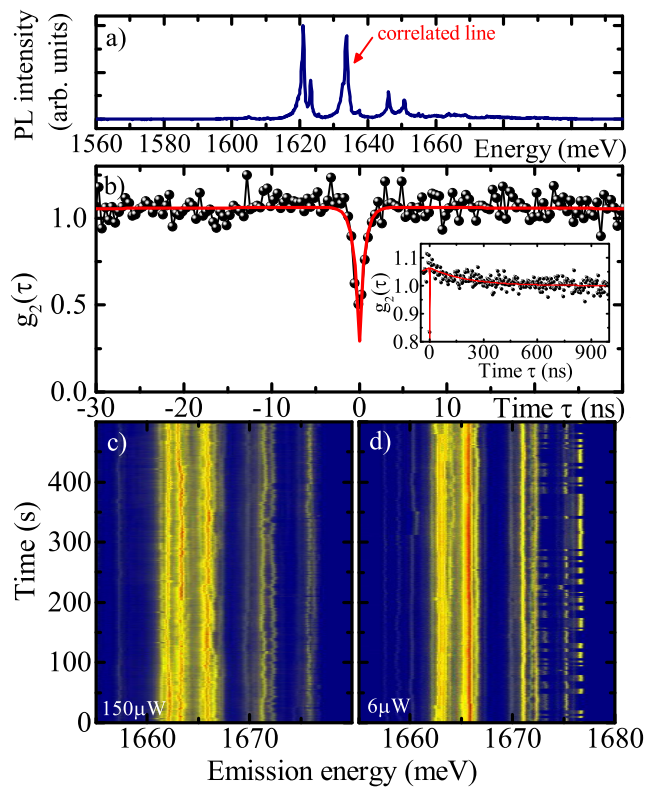}
\caption{\textbf{Narrow line PL centers as sources of single
photon emission.} \textbf{a}, Narrow line emission spectrum
detected at the micro-spot located at the edge of one of the thick
WSe$_2$ flakes. \textbf{b}, Photon coincidence correlation,
g$_2$($\tau$), which displays the effect of photon antibunching,
with a characteristic time of $\sim$600ps, for one of the narrow
emission lines marked in \textbf{a}. A weak bunching effect on a
longer time scale (200 ns), shown in the inset, is likely caused
by temporal jittering of the center of the line. \textbf{c}, Two
sets of narrow line emission spectra, recorded one after another,
with 100 ms resolution, measured under two different powers of
laser excitations (150 and 6 $\mu$W), which demonstrate the line
jittering effects being less pronounced at lower excitation
powers.} \label{fig:cor}
\end{figure}

Single isolated centers in solids often act as efficient,
high-fidelity single photon emitters~\cite{lounis2005}. The NLECs
at the edges of WSe$_2$ flakes exhibit this property as well. This
is illustrated in Fig.~\ref{fig:cor} with results of single photon
correlation measurements performed on one of the selected sharp
emission lines. Photon antibunching in the autocorrelation
function is clear. The characteristic coincidence time in the
auto-correlation function, which yields the upper bound for the
lifetime of the emitting state, is estimated to be about $600$~ps
for this particular line and reaches up to a few nanoseconds for
other investigated lines. The NLECs at the edges of WSe$_2$ flakes
appear to be overall robust on a long time scale and survive many
temperature cycless (room to helium temperature) including
prolonged exposures to ambient (air) conditions. Nevertheless, our
NLECs show clear fluctuation effects on a short time scale:
jittering of centers of lines, of the order of the linewidth, on a
millisecond time scale as well as larger jumps of lines on a time
scale of seconds (see Fig.~\ref{fig:cor}c). Stability is improved
when decreasing the power of the laser excitation. This behavior
is characteristic of many other single photon emitters, such as,
for example semiconductor quantum dots, and it is commonly
associated with fluctuations of electric charge in the surrounding
of the emitting center~\cite{pietka2013}.

\begin{figure}
\includegraphics[width=8.5cm]{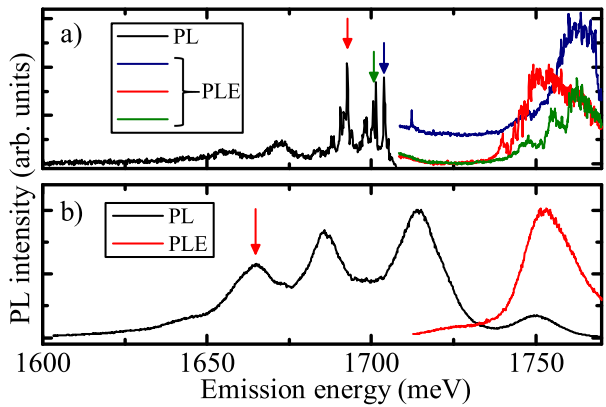}
\caption{\textbf{Absorption resonances of narrow-line-emission
centers and of a 2D WSe$_2$ monolayer.} PL excitation spectra
measured in the range available with a Ti:Sapphire laser for,
\textbf{a}, three selected narrow PL lines associated with the
edge of the thick WSe$_2$ flake, and of, \textbf{b}, the localized
exciton band of the 2D WSe$_2$ monolayer. The detection energies
are shown with arrows. The approximately coinciding absorption
resonances underline a link between the electronic states of the
edge emission centers and of the WSe$_2$ monolayer.}
\label{fig:ple}
\end{figure}

\begin{figure}
\includegraphics[width=8.5cm]{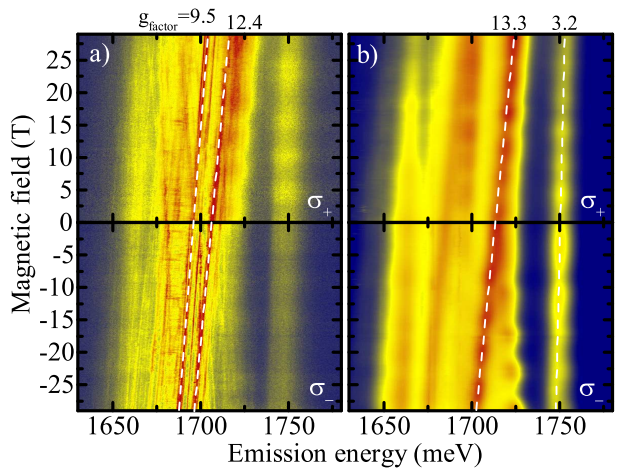}
\caption{\textbf{Narrow line emission centers and a 2D WSe$_2$
monolayer in a magnetic field.} Intensity contour plots of
micro-magneto-PL spectra measured in the Faraday configuration:
\textbf{a}, from the edge spot giving rise to sharp emission lines
and, \textbf{b}, from the center of the WSe$_2$ monolayer. The top
and bottom panels correspond to the spectra, resolved in circular
polarizations $\sigma_+$ and $\sigma_-$, respectively. The
characteristic amplitude of the splitting of the emission lines
into $\sigma_+$ and $\sigma_-$ components (see white dashed white
traces) is expressed in terms of g-factors (splitting =
g$\mu_B$B). Note, the anomalously large but, at the same time,
similar splitting for the narrow lines and for one of the broad PL
peaks (due to a charged exciton) of the monolayer. The
oscillations of the PL intensity with the magnetic field visible
in the spectra are due to the use of fiber optics in our
experiments (see Methods).} \label{fig:mag}
\end{figure}

Having established the single-photon-emitter character of NLECs at
the edges of WSe$_2$ flakes, we now focus more on their optical
properties. The first task is to identify their absorption
resonances. The PL-excitation spectra of three selected narrow
lines (of the NLEC at the edge of the thick WSe$_2$ flake) are
presented in Fig.~\ref{fig:ple}a. Each of these lines resonates in
a slightly different energy range, though all three absorption
resonances overlap quite well with the characteristic absorption
band of the WSe$_2$ monolayer film (see Fig.~\ref{fig:ple}b). This
approximate coincidence highlights the link between the electronic
band structures of our NLECs and of the WSe$_2$ monolayer; the
statement being also applied to the case when NLECs are attached
to thick WSe$_2$ films.

The fact that electronic structure of our NLECs may originate from
that of the WSe$_2$ monolayer is further supported by the
magneto-PL measurements. These experiments have been carried out
in a large range of magnetic fields (up to 29T), in the Faraday
configuration, and $\sigma_+$ and $\sigma_{-}$ circular
polarization components of the emitted light were resolved (by
inverting the direction of the magnetic field while passing the
emitted light through a fixed ensemble of a linear polarizer and a
$\lambda$/4 plate). In Fig.~\ref{fig:mag}, we compare the
magneto-PL response of the WSe$_2$ monolayer film with that of the
NLEC attached to this layer. The details of these data remain to
be thoroughly analyzed. Here we focus on one of their prominent
features, relevant for this work. Remarkably, we observe a
striking resemblance in the amplitude of the
$\sigma_{+}$/$\sigma_{-}$ splitting (Zeeman splitting) for the
charged exciton of the two-dimensional monolayer and for the
narrow emission lines, in that both show an anomalously large
Zeeman effect. The characteristic amplitude of the g-factor
(splitting = $g \mu_B B$) for charged excitons is g$\sim$13 and
similarly large values (g$\sim9 - 12$) are found for the emission
lines of our NLECs. The difference in the amplitude of the Zeeman
splitting for the neutral (g$\sim$3.2) and charged exciton of the
WSe$_2$ monolayer remains a puzzle as already reported in
Ref.~\onlinecite{srivastava2014}. An overall large Zeeman effect
in WSe$_2$ is likely due to a strong, spin-orbit interaction in
this material. This issue remains to be clarified, but the
similarity between Zeeman effects found for the WSe$_2$ monolayer
and for the NLECs is an additional indication that the matrix of
our NLECs might be formed out of the WSe$_2$ monolayer.

A firm identification of NLECs seen at the edges of the WSe$_2$
flakes is an obvious challenge. Our working hypothesis is that
these NLECs consist of nano-sized, lateral fragments (nano-flakes)
of a WSe$_2$ monolayer, which are apparently formed at the edges
of exfoliated flakes of monolayers as well as WSe$_2$ multilayers.
Notably, the appearance of monolayer nano-flakes at the edges of
thick multilayers may not be surprising as it is quite
characteristic for the exfoliation technique that even large
monolayers can be found on the sides of thick flakes of various
exfoliated/transferred materials (e.g. graphene)~\cite{blake2007}.
Nonetheless, these nano-flakes are usually isolated and display
relevant quantum confinement effects, what accounts for the
observation of narrow and single-photon emission lines. As such,
our NLECs can be seen as quantum dots made of the WSe$_2$
monolayer, objects similar to those extensively studies in
conventional semiconductor structures~\cite{yamamoto2005}. The
resemblance of the optical responses of the WSe$_2$ monolayer and
the NLECs (similar emission spectral range, similar absorption
resonances and Zeeman splittings) favors these claims. Since the
spectral range of sharp emission line covers the PL band of trions
and/or bound exciton of the WSe$_2$ monolayer, the nano-flakes
should act as efficient traps for the electric charge; the
observed sharp emission lines must then be due to charged
excitonic complexes as well. The characteristic,
temperature-activated broadening of NLEC spectra (not shown)
resembles effects found in semiconductor quantum dots (acoustic
phonon broadening~\cite{besombes2001}) and additionally supports
the analogies invoked above.

The monolayer nano-flake (quantum dot) scenario for our NLECs is
plausible but obviously needs further clarifications/work and
other hypotheses should be discussed at present. In this light,
our unsuccessful efforts to observe any signature of a sequential
(cascaded) emission in photon cross-correlation measurements
between any pair of the NLEC lines tested are disappointing from
the point of view of the quantum dot scenario. This sequential
emission, a prominent example of which is the biexciton~-~exciton
cascade, is a distinct feature of semiconductor quantum dots (and
nanocrystals)~\cite{moreau2001}. On the other hand, this feature
is absent for all other single photon-emitters, such as organic
molecules, colored centers in crystals~\cite{lounis2005} and also
individual carbon nanotubes~\cite{hogele2008}. The possibility
that our NLECs are due to an insufficiently clean exfoliation
procedures, as a result of which some organic molecules are
attached and functionalized at the edges of WSe$2$ flakes, is
unlikely. This is because of the overall robustness of our NLECs
and the anomalously large Zeeman effect (unexpected for organic,
carbon based molecules with a small spin-orbit interaction). So
far, all our efforts to localize the NLECs with AFM measurements
have failed, but this is perhaps not surprising due to the
insufficient spatial resolution of this technique. Instead,
however, we have systematically observed that the thickness of our
WSe$_2$ films is slightly enhanced at the side of the flake. This
may also signify the effect of curling of the WSe$_2$ layer at the
flake edges~\cite{huang2009}, but suggesting the edge defects in
form of a WSe$_2$ nanotube~\cite{natha2001} is highly speculative.
Nevertheless, the latter, as well as any other defect evoking
hypothesis, is to be tested in the future. This, we believe, can
be efficiently done with scanning tunneling
microscopy/spectroscopy as our NLECs appear on an open surface and
WSe$_2$ flakes can be suitably transferred onto a conducting
(e.~g.~graphene) substrates.

Concluding, we have identified single photon emitting centers
which are located at the edges of WSe$2$ flakes (mono- and
multi-layers) exfoliated onto Si/SiO$_2$ substrates. The
characteristic spectra of these centers, seen at low temperatures
($\leq20$~K), are composed of a series of sharp emission lines
which clearly reveal the effect of photon-antibunching. In a
number of aspects, the optical response of the sharp emission
lines resembles that of the WSe$_2$ monolayer: overlapping
spectral emission range, coinciding absorption resonances and
similar anomalously large Zeeman splittings. With these
observations, the single photon emitters reported here are
recognized as nano-flakes (quantum dots) of the WSe$_2$ monolayer,
located at the edges of conventionally studied WSe$_2$ films. Our
findings may open a new field of quantum optoelectronics studies
of thin layers of semiconducting transition metal dichalcogenides.
When setting up this manuscript, we have noticed another
paper~\cite{srivastava2014_2} reporting findings similar to ours.
\vspace{-0.5cm}
\section*{Methods}
\vspace{-0.2cm}
The samples used in the experiments were prepared by mechanical
exfoliation of bulk WSe$_{2}$. We followed the procedure which has
recently been demonstrated in Ref.~\onlinecite{gomez2014}. For
thinning down the flakes we used F07 acrylic tape from Microworld
and for transferring them onto a target substrate (a piece of
undoped silicon wafer with an 86\hspace{0.5mm} nm-thick layer of
thermally grown silicon dioxide on top),
polydimethylosiloxane-based elastomeric films from Gel-Pak. All
Si/SiO$_{2}$ substrates had been photolithographically equipped
with Ti/Au alignment markers to facilitate locating the flakes
under different optical setups and ashed with oxygen plasma
shortly before exfoliation to clean and activate their surface.
The AFM characterization of the flakes was performed in a tapping
mode with the aid of a Veeco Dimension 3100 microscope.

The optical studies were carried-out in $\mu$PL setups. The
diameter of the laser beam was typically about 1~$\mu$m. The
spectra were resolved with a 50 cm monochromator with a $600$ or
$1800$~g/mm grating and detected with a charge coupled device
(CCD) camera. The magneto-optical studies in the Faraday
configuration were performed in a resistive magnet supplying
magnetic fields up to 29T. In this setup the sample was mounted on
an x-y-z piezo-stage allowing the positioning of the sample with
submicrometer precision. Optical fibers were employed in the
set-up, to transfer the excitation light (from tunable Ti:Sapphire
laser) and to collect the PL signal. The $\mu$PL setup was placed
in a probe, filled with helium exchange gas and cooled down to 4.2
K. A $\lambda$/4 waveplate and a linear polarizer were aligned as
a circular polarizer in front of the entrance of the detection
fiber, in order to obtain circular polarization resolution. Due to
the presence of optical fibers, the magnetic-field induced
rotation of the linear polarization angle (Faraday effect)
introduced oscillations of the emission intensity observed in the
magnetic field evolution of the PL spectrum.

Photon correlation measurements were performed in a Hanbury-Brown
and Twiss configuration with an argon laser ($488.0$~nm)
excitation. Avalanche photodiodes were used for photon detection,
with a temporal resolution of about 300 ps. The sample resided in
a flow cryostat at a temperature of 4.2K. The correlation function
obtained in the experiments was described with the following
formula:

\begin{equation}
g_2 \left( \tau \right) = 1-A_1*\exp \left( -\left| \frac{\tau}{t_1} \right| \right)+A_2*\exp \left( \left| \frac{\tau}{t_2} \right| \right)
\end{equation}
where $t_1$ is a characteristic time for the photon anti-bunching,
$t_2$ is a characteristic time of the long-timescale bunching,
$A_1$ is the antibunching depth and $A_2$ is the bunching
amplitude. In case of the example correlation function shown here,
the depth of the antibunching was equal to $A_1=0.77 \pm 0.08$,
which signifies that the emitter is a single photon source. The
characteristic time was $t_1=0.61 \pm 0.08$ns. The time of the
long-time bunching was $t_2=204 \pm 22$ns with a small amplitude
$A_2=0.064 \pm 0.004$.
\section*{Acknowledgments}
The authors thank M.L. Sadowski for valuable discussions and I. Breslavetz for technical assistance,  and
acknowledge the support from the European Research Council (MOMB
project No. 320590), the EC Graphene Flagship project (No. 604391)
and the NCN "Harmonia" programme.

\end{document}